\documentclass[12pt]{article}

\usepackage{amsmath}
\usepackage{amssymb}
\usepackage{graphicx}
\newcommand{\etal}{{\it et al}}

\begin{document}

\title{Berezinskii-Kosterlitz-Thouless transitions in the six-state clock
model}

\author{Haruhiko Matsuo and Kiyohide Nomura \\
Department of Physics, Kyushu University \\
Fukuoka 812-8581 JAPAN \\
{\tt halm@stat.phys.kyushu-u.ac.jp}\\
{\tt knomura@stat.phys.kyushu-u.ac.jp}}


\maketitle

\begin{abstract}
Classical 2D clock model is known to have a critical phase with
Berezinskii-Kosterlitz-Thouless(BKT) transitions. These transitions have
logarithmic corrections which make numerical analysis difficult. In
order to resolve this difficulty, one of the authors has proposed the
method called level spectroscopy, which is based on the conformal field
theory. We extend this method to the multi-degenerated case. As an
example, we study the classical 2D 6-clock model which can be mapped to
the quantum self-dual 1D 6-clock model. Additionally, we confirm that
the self-dual point has a precise numerical agreement with the
analytical result, and we argue the degeneracy of the excitation states
at the self-dual point from the effective field theoretical point of
view.
\end{abstract}

\section{Introduction}
Although low dimensional spin systems with a continuous rotational
symmetry, like the 2D XY spin model, don't have a symmetry breaking phase
at finite temperatures, it doesn't mean the absence of phase transitions.
Actually Berezinskii\cite{berezinskii70:_destr} and Kosterlitz and
Thouless\cite{j.m.kosterlitz73:_order} pointed out the vortex-antivortex
excitation phase transition in the 2D XY model, which is known as the BKT
transition. 

The classical 2D clock model, which has a discrete rotational symmetry,
is also known to have a BKT critical region. Jos\'e
\etal\cite{j.v.jose77:_renor} studied the 2D classical $p$-state clock
model by the renormalization group analysis and pointed out the presence
of the intermediate critical phase for large $p$.  Elitzur
\etal\cite{s.79:_phase_abelian} have also obtained the same result
through the correlation inequality argument.

We map the 2D $p$-clock model onto its quantum 1D $p$-clock model and
analyze the latter. This is based on the fact that the $(1+d)$
dimensional classical statistical mechanics can be mapped onto the
$d$-dimensional quantum statistical mechanics. Here, we should be
careful that the 2D classical clock model, introduced the following
section, doesn't have the self-duality {\it explicitly}, although its
quantized Hamiltonian has the self-duality. It may seems strange, but
thinking of the universality class; that is, the $Z_p$ Villain model and
the 2D classical $p$-clock model are considered belong to the same
universality class, the self-duality of the quantum clock model is
acceptable. Additionally, the corresponding continuum model(the dual
sine-Gordon model) also has the self-duality. We will see an excellent
numerical agreement with the self-duality.

Generally, the BKT transition points were difficult to determine
numerically because of the logarithmic corrections and the anomalous
divergence of the correlation length. But the level spectroscopy
method\cite{nomura94:_critic_s_xxz}\cite{nomura95:_correl_gordon} has
been proposed to resolve these problems. Near a BKT transition point,
some scaling dimensions change from relevant to irrelevant or vice
versa. Choosing some applicable scaling dimensions, we can determine the
BKT transition point by the crossing point of these scaling
dimensions. In addition, we can eliminate logarithmic corrections
accompanying the BKT transition. This is the brief explanation of the
level spectroscopy.

So far the level spectroscopy has been applied to the three cases.  The
first case is between the BKT critical phase and the phase of the
non-degenerate ground
state\cite{nomura95:_correl_gordon}\cite{nomura98:_su_z_bkt}. The second
is between the BKT critical phase and the phase of 2-fold degenerate
ground state\cite{nomura94:_critic_s_xxz}. The third is between the BKT
critical phase and the multi-degenerated ground state phase. Recently
Tonegawa \etal\cite{tonegawa04:_magnet_s} studied the $1/3$
plateau problem of the $S=\frac{1}{2}$ antiferromagnetic XXZ chain with
the $Z_3$ symmetry breaking and Otsuka \etal\cite{otsuka05}
studied the 2D AF $3$-state Potts model with the $Z_6$ symmetry breaking
case.

In this paper, we apply the level spectroscopy to the 1D quantum
$6$-clock model with duality and use the duality relation to check the
level spectroscopy results between the BKT critical phase and the
multi-degenerated ground state from another point of view than former
studies. In addition, we discuss the degeneracy at the self-dual point
from the field theoretical point of view. The self-dual point of this
model is trivial but it will be important in the 2D AF $3$-state Potts
model which have no explicit self-duality.

\section{Theory}
\subsection{Discrete model}
In this subsection, we review the 1D quantum $p$-state clock model with
a nearest neighbor interaction as a Hamiltonian limit of a 2D classical
$p$-state clock model on a square lattice. The 2D classical Hamiltonian is
\begin{eqnarray}
 \mathcal{H} &=& -\beta \sum_{\langle i,j \rangle} 
  \{ J_s \left[\cos\left(\Theta_{i,j} - \Theta_{i+1,j}\right)
  -1\right] \nonumber\\
  &&      + J_{\tau} [\cos (\Theta_{i,j} - \Theta_{i,j+1}  ) ] \},
\end{eqnarray}
where $\Theta_{i,j} = (2\pi r/p), \,r=0,1,\ldots,p-1$ is the clock spin
variable. When we take the Hamiltonian limit $\beta J_s\rightarrow 0$,
$\beta J_{\tau} \rightarrow\infty$ with $\lambda$
fixed\cite{solyom81:_renor_hamil_potts}, we obtain the 1D quantum
Hamiltonian \cite{s.79:_phase_abelian}\cite{solyom81:_renor_hamil_potts}
\begin{eqnarray}
\label{eq:hamil0}
 H = -2\sum_{n=1}^N \left\{ \lambda \cos({\hat\Theta}_n -
		     {\hat\Theta}_{n+1}) + \cos {\hat p}_n \right\}
\end{eqnarray}
where $\lambda$ is the transverse field of the system. It can be
interpreted as the inverse of the temperature.

When we define the operators $\sigma_n,\Gamma_n$ as 
$\sigma_n = \exp i{\hat\Theta}, \Gamma_n = \exp i{\hat p}$, 
we obtain another representation of the $1$D quantum Hamiltonian
(\ref{eq:hamil0}) as
\begin{eqnarray}
 H = -\sum_{n=1}^{L} \left\{ \lambda\left( \sigma_n\sigma^+_{n+1} 
     + \sigma^+_n\sigma_{n+1}\right) + \Gamma_n + \Gamma^+_n\right\},
 \label{eq:hamil1}
\end{eqnarray}
where $\omega = \exp(2\pi i/p)$, and
$$ \sigma_n =
\left(
\begin{array}{ccccc}
 1& & & & \\
  & \omega & & & \\
 & & \omega^2 & & \\
 & & & \ddots & \\
 & & & & \omega^{p-1}
\end{array}
\right), \quad
\Gamma_n = 
\left(
\begin{array}{ccccc}
 0& & & & 1\\
 1 & 0& & & \\
 & 1& 0& & \\
 & & \ddots & \ddots& \\
 & & & 1 & 0
\end{array}
\right).
$$ 
These operators satisfy $(\sigma_n)^p = (\Gamma_n)^p = 1$.

This model has a $Z_p$ charge symmetry. The $Z_p$ charge operator is 
defined as
\begin{eqnarray}
 U_Q = \prod_{n=1}^{L} \Gamma_{n}.
\end{eqnarray}
which commutes with the Hamiltonian $(\ref{eq:hamil1})$. $U_Q$ has
eigenvalues $\omega^Q, Q=0,1,\cdots,p-1$. Because of the conservation, 
$U_Q$ splits $H$ into the charge sectors 
whose corresponding eigenspaces are labeled by $Q=0, 1, \cdots,p-1$,
i.e. $H$ is able to be block-diagonalized. 
Also we can block-diagonalize it with the toroidal boundary conditions:
\begin{eqnarray}
 \label{eq:toroidal}
 \sigma_{N+1} = \exp \left( \frac{2\pi i}{p} {\tilde Q} \right) \sigma_1,
\,{\tilde Q} = 0, 1, \ldots , p-1.
\end{eqnarray}
${\tilde Q} = 0$ is the periodic boundary condition and ${\tilde Q} =
p/2$ ($p$ even) is the twisted boundary condition. Hereafter we will
assume that $p$ is even.  Implementing the twisted boundary condition to
the system corresponds to adding a half-charge to the sine-Gordon
model\cite{nomura98:_su_z_bkt}
\cite{destri89:_twist_bound_condit_in_confor_invar_theor}
which is the continuum limit of the discrete model.

There exits a unitary operator such that $\sigma_n$ is transformed into
$\Gamma^+_n$ and also $\Gamma_n$ is transformed into $\sigma_n$. For example
we can take its matrix elements as
\begin{eqnarray}
 U_{i,j} = \frac{1}{\sqrt{p}} \omega^{(1-i)(1-j)}.
\end{eqnarray}
This operator $U$ gives the relation
\begin{eqnarray}
 U \sigma_n U^+ &=& \Gamma_n^+, \\
 U \Gamma_n U^+ &=& \sigma_n.
\end{eqnarray}
By this relation, we get another representation of the 1D quantum
Hamiltonian as 
\begin{eqnarray}
  H' = - \sum_{n=1}^{L} \left\{\lambda
         \left( \Gamma_n\Gamma^+_{n+1} + \Gamma^+_n\Gamma_{n+1}\right)
         + \sigma_n + \sigma^+_n \right\}. \label{eq:hamil2}
\end{eqnarray}
In this case, notice that the $Z_p$ charge operator should be
\begin{eqnarray}
 U'_Q = \prod_{n=1}^L \sigma_n. \label{eq:unitary}
\end{eqnarray}
The form $(\ref{eq:hamil2}), (\ref{eq:unitary})$ is more useful for
numerical calculation than the pribious form $(\ref{eq:hamil1})$, when
we choose the $\sigma_n$ diagonal representation.

There is another non-local transformation between $\sigma$ and $\Gamma$:
\begin{eqnarray}
\label{eq:dualtrans}
 \sigma_n &\rightarrow& {\tilde\sigma}_n = \Gamma_{n+1}\Gamma^+_n 
  \quad\mathrm{for}\, n<N \\ \nonumber
 \sigma_N &\rightarrow& {\tilde\sigma}_N = \Gamma^+_N \\ \nonumber
 \Gamma_n &\rightarrow& {\tilde\Gamma}_n = \prod_{l=1}^n \sigma_l.
\end{eqnarray}
Applying this transformation to the Hamiltonian $(\ref{eq:hamil2})$ and
the help of 
\begin{eqnarray}
{\tilde\Gamma}_n {\tilde\Gamma}^+_{n+1} 
= \prod_{l=1}^n \sigma_l \prod_{l=1}^{n+1} \sigma^+_l = \sigma^+_{n+1},
\end{eqnarray}
we get the dual form of $(\ref{eq:hamil2})$ as
\begin{eqnarray}
-\sum_{n=1}^{L} \left[ \lambda (\sigma^+_n + \sigma_n)
			     + \Gamma_n\Gamma^+_{n+1}
			     + \Gamma^+_n\Gamma_{n+1}\right].
\end{eqnarray}
Therefore the 1D $p$-state quantum clock model satisfies the following
duality relation,
\begin{eqnarray}
\label{eq:self_duality}
 H(\lambda) = \lambda H(1/\lambda). 
\end{eqnarray}
The self-dual point is $\lambda_c = 1$.

The charge conjugation operator is defined as 
\begin{eqnarray}
C= \prod_{n=1}^L c_n, \quad
 c_n = \left(
\begin{array}{ccccc}
 1 & & & & 0 \\
   & 0 & & 0 &1 \\
   &   &  \ddots & 1  & \\
 & 0 & 1 & \ddots & \\
0 & 1 & & & 0
\end{array}
\right),
\end{eqnarray}
which has eigenvalues $\pm1$ and transforms $\sigma_n$ and $\Gamma_n$ as
\begin{eqnarray}
 c_n \sigma_n c_n^+ &=& \sigma^+_n \\
 c_n \Gamma_n c_n^+ &=& \Gamma^+_n.
\end{eqnarray}
By this relation, it is easy to see that the
Hamiltonian (\ref{eq:hamil1}) and (\ref{eq:hamil2}) commutes with $C$.

$Z_p$ charge operator $U_Q$ satisfies
\begin{eqnarray}
 C U_Q C = U^+_Q.
\end{eqnarray}
So the eigenvalues $\omega^Q$ of $U_Q$ become $\omega^{-Q}$, that is,
generally the eigenstates of $U_Q$ are not those of $C$.  But the
eigenstates of $Q=0,p/2$ are invariant under this transformation,
therefore these eigenstates have eigenvalues $C=\pm 1$. We will use this
fact to classify the eigenstates.

\subsection{Effective model\label{effective}}
As an effective theory of a 1D quantum spin system, the sine-Gordon
model has been studied.
Here, we will explain the $Z_p$ dual
sine-Gordon model\cite{wiegmann78:_one_fermi},
\begin{eqnarray}
 \mathcal{L} = \frac{1}{2\pi K} (\nabla \phi)^2
  + \frac{y_{\phi}}{2\pi\alpha^2}\cos \sqrt{2}\phi
  + \frac{y_{\theta}}{2\pi\alpha^2}\cos p\sqrt{2}\theta,
  \label{sGeq1}
\end{eqnarray}
where $K$ is also inverse of temperature as $\lambda$, $\alpha$ is
an ultraviolet cutoff and
$\phi$ and $\theta$ are periodic:
\begin{eqnarray}
 \phi &=&  \phi + \frac{2\pi}{\sqrt{2}}, \\
 \theta &=& \theta + \frac{2\pi}{\sqrt{2}}.
\end{eqnarray}
The field $\theta$ is dual to $\phi$ and they satisfy the relation
\begin{eqnarray}
 \partial_x \phi = - \partial_y (K\theta), \quad \partial_y \phi
  = \partial_x (K\theta). \label{eq:duality}
\end{eqnarray}
This model is the effective model for the Hamiltonian $(\ref{eq:hamil1}),
(\ref{eq:hamil2})$.

 The vertex operators are defined as
\begin{eqnarray}
 O_{m,n} = \exp(im \sqrt{2}\theta) \exp(in\sqrt{2}\phi).
  \label{eq:vertexOp1}
\end{eqnarray}
When $y_{\phi} = y_{\theta} = 0$, its scaling dimension is
\begin{eqnarray}
 x_{m,n}(K) =  \frac{1}{2} \left( n^2 K + \frac{m^2}{K}\right).
\end{eqnarray}

In a high temperature region ($K$ small) the second term of
$(\ref{sGeq1})$ is more relevant than the third term. So the
renormalization group behaviour in the high temperature region can be studied
through the Lagrangian
\begin{eqnarray}
 \mathcal{L}_{\rm high} = \frac{1}{2\pi K} (\nabla \phi)^2
    + \frac{y_{\phi}}{2\pi\alpha^2}\cos \sqrt{2}\phi
    \label{sGeq2}.
\end{eqnarray}
When $K=4$ this sine-Gordon model shows the BKT transition, and that is
studied by the level spectroscopy\cite{nomura98:_su_z_bkt}.  On the
other hand, in the low temperature region the second term of $(\ref{sGeq1})$
is more irrelevant than the third term.  Then we deal with the
sine-Gordon Lagrangian
\begin{eqnarray}
 \mathcal{L}_{\rm low} = \frac{K}{2\pi} (\nabla \theta)^2
    + \frac{y_{\theta}}{2\pi\alpha^2}\cos p\sqrt{2}\theta \label{sGeq3}
\end{eqnarray}
with the $Z_p$ symmetry breaking term. Here we got the first term through
substituting $(\ref{sGeq1})$ to $(\ref{eq:duality})$.

Under the change of the cutoff $\alpha \rightarrow e^l\alpha$ and
parameterizing $K / 4 = 1 + y_0 / 2$, the
renormalization group equations for $(\ref{sGeq2})$ are
\begin{eqnarray}
 \frac{d y_0(l)}{dl} = -y_{\phi}^2(l), \quad \frac{dy_{\phi}(l)}{dl} = -y_{\phi}(l)y_0(l)
\end{eqnarray}
where $l$ is related to $e^l = L$ for the finite system size.  For the case
$(\ref{sGeq3})$, by parameterizing $K^{-1} p^2 / 4 = 1 +y_0/ 2$ and using
$y_{\theta}$ instead of $y_{\phi}$, we get
the renormalization equations
\begin{eqnarray}
 \frac{d y_0(l)}{dl} = -y_{\theta}^2(l), \quad \frac{dy_{\theta}(l)}{dl} = -y_{\theta}(l)y_0(l)
\end{eqnarray}

In the high temperature limit $\lambda \rightarrow 0$ (or $K \rightarrow 0$)
this model has a non-degenerated ground state. The transition point
between the non-degenerated ground state phase and the BKT critical
phase can be determined by the $K=4$ version of the level
spectroscopy. In this case, the renormalized scaling dimensions are
given in \cite{nomura98:_su_z_bkt}.
In the low temperature limit $\lambda \rightarrow \infty$, the model has
a $p$-fold degenerated ground state. In this case, the simple $K=4$
version of the level spectroscopy is no longer available. The BKT
critical transition point can be determined by the $K=p^2/4$ version of
the level spectroscopy (see also \cite{nomura95:_correl_gordon} and
\cite{otsuka05}).  The renormalized scaling dimensions of near $K=p^2/4$
are shown in Table\ref{tab:scaling_ops}(note that implementing TBC is
the same as considering the half-charge of the effective model
\cite{destri89:_twist_bound_condit_in_confor_invar_theor}).
\begin{table}
 \caption{Operator content of the $K=p^2/4$ sine-Gordon model with $Z_p$
 symmetry breaking term. TBC is short for twisted boundary
 condition. Others are under periodic boundary condition.}
 \label{tab:scaling_ops}
  \begin{tabular}{cccccc}
\hline
   field & \multicolumn{2}{c}{quantum number}&  B.C. & scaling   & renormalized\\
   &         &      & & dimension & scaling \\
   & $Q$     & $C$  & & $x_{m,n}$ & dimension\\
\hline
   $\exp (im\sqrt{2}\theta)$  & $m \ne p/2,p $ & &  PBC  & $x_{m,0}$ & $\frac{2m^2}{p^2}\left(1+\frac{y_0}{2}\right)$\\[3pt]
   $\cos (\sqrt{2}p\theta/2)$    &    $p/2$ & $1$ & PBC & $x_{p/2,0}$ & $\frac{1}{2}\left(1-\frac{3}{2}y_0\right)$ \\[3pt]
   $\sin (\sqrt{2}p\theta/2)$    &    $p/2$ & $-1$ & PBC & $x_{p/2,0}$ & $\frac{1}{2}\left(1+\frac{1}{2}y_0\right)$ \\[3pt]
   marginal & $p$ & $1$& PBC &$x_{p,0}$ & \\[3pt]
   &    $0$ & $1$ & PBC & $x_{0,1}$ & $\frac{9}{2} \left(1-\frac{1}{2}y_0\right)$ \\[3pt]
   $\cos(\sqrt{2}\phi/2)$ &   $0$ &  $1$ &  TBC & $x_{0,\frac{1}{2}}$ & $\frac{9}{8}\left(1-\frac{1}{2}y_0\right)$ \\[3pt]
   $\sin(\sqrt{2}\phi/2)$ &   $0$ & $-1$ &  TBC & $x_{0,\frac{1}{2}}$ & \\
\hline
  \end{tabular}
\end{table}

 Assuming conformal invariance, the scaling dimensions $x_{m,n}$ are
 related to the energy gap of the finite size system with periodic
 boundary conditions as
 \begin{eqnarray}
  x_{m,n} = \frac{L}{2\pi v} \left( E_{m,n}(L) - E_{\rm g}(L)\right),
 \end{eqnarray}
 where $L$ is the system size, $E_{\rm g}(L)$ is the ground state
 energy, and $v$ is the velocity of the
 system\cite{cardy86:_operat_conten_of_two_dimen}. And the central
 charge $c$, which we will use to confirm the universality class of the
 system, is given by the finite size correction of the system as
 \begin{eqnarray}
  E_{\rm g}(L) = e_{\rm g} L - \frac{\pi vc}{6L}
 \end{eqnarray}
 where $e_{\rm g}$ is the bulk ground state energy per
 site\cite{affleck85:_univer_term_free_energ_critic}.

 At last, we discuss the dual transformation in this dual sine-Gordon
 model, corresponding with the discrete case
 $(\ref{eq:dualtrans}),(\ref{eq:self_duality})$.  When we apply the dual
 transformation
   \begin{eqnarray}
    \phi \rightarrow p\theta, \quad p\theta \rightarrow \phi\label{sG_dual}
   \end{eqnarray}
   to the dual sine-Gordon model $(\ref{sGeq1})$ , we get
   the dual form
   \begin{eqnarray}
    \mathcal{L}_{\rm dual} &=& \frac{p^2}{2\pi K}
     \left(\nabla\theta\right)^2 + 
     \frac{y_{\phi}}{2\pi \alpha^2} \cos p\sqrt{2}\theta
     + \frac{y_{\theta}}{2\pi\alpha^2} \cos\sqrt{2}\phi.
     \label{eq:dual_sineGordon}
   \end{eqnarray}
   When we substitute ($\ref{eq:duality}$) to
   ($\ref{eq:dual_sineGordon}$), we obtain the same form as
   $(\ref{sGeq1})$ aside the coefficients; assuming $y_{\phi} =
   y_{\theta}$ and $p=K$, including the coefficients, we get the same
   equation as $(\ref{sGeq1})$. This is the self-dual point of the dual
   sine-Gordon model. Note that with the dual transformation
   $(\ref{sG_dual})$, fields in Table \ref{tab:scaling_ops} are
   exchanged as $\cos \sqrt{2} p\theta/2 \leftrightarrow \cos \sqrt{2}
   p/2$ etc.  These exchanges mean the degeneracy of the excitation
   spectra at the self-dual point.

   The quantum discrete model has the self-dual
   point$(\ref{eq:self_duality})$, and now this self-duality is extended
   to the continuum model. In the following section, we numerically
   estimate the self-dual point with good precision.

\section{Numerical Results}
 In the previous section we have argued the general $p$-state clock
 model. In this section we show the numerical results for the $p=6$
 case. The ground states with the various $Z_p$ charges and the boundary
 conditions are obtained by the Lanczos method. The system sizes are up
 to $L=10$.

 First, we determine the two BKT transition points. These points have
 essentially different property, because one is located between the
 non-degenerate ground state phase and the BKT critical phase, the
 another is located between the 6-fold degenerate ground states phase
 and the BKT critical phase.
 
 Next, to confirm the universality class of the BKT critical phase,
 we calculate the ratio of the scaling dimensions,
 the central charge $c$ and the parameter $K$ of the sine-Gordon model.
 
  \subsection{The BKT transition point in lower $\lambda$}
  The BKT transition point between the non-degenerate ground state phase
  and the BKT critical phase is given by the level crossing of the
  following energy gaps\cite{nomura98:_su_z_bkt}:
  \begin{eqnarray}
   x_{2,0} &=& \frac{L}{2\pi v} \Delta E(Q=2, {\rm PBC}),\\[8pt]
   x_{0,\frac{1}{2}} &=& \frac{L}{2\pi v} \Delta E(Q=0, {\rm TBC}, C=1).
  \end{eqnarray}
  $\Delta E$ is an energy gap between the ground state and the excitation
  state with the quantum number $Q$ and a periodic boundary
  condition(PBC) or a twisted boundary condition(TBC). In terms of the
  toroidal boundary condition$(\ref{eq:toroidal})$, PBC corresponds to
  ${\tilde Q} = 0$, and TBC corresponds to ${\tilde Q} = p/2$. The
  ground state has $Q=0, C=1$ with PBC.

\begin{figure}
   \includegraphics[scale=1.0]{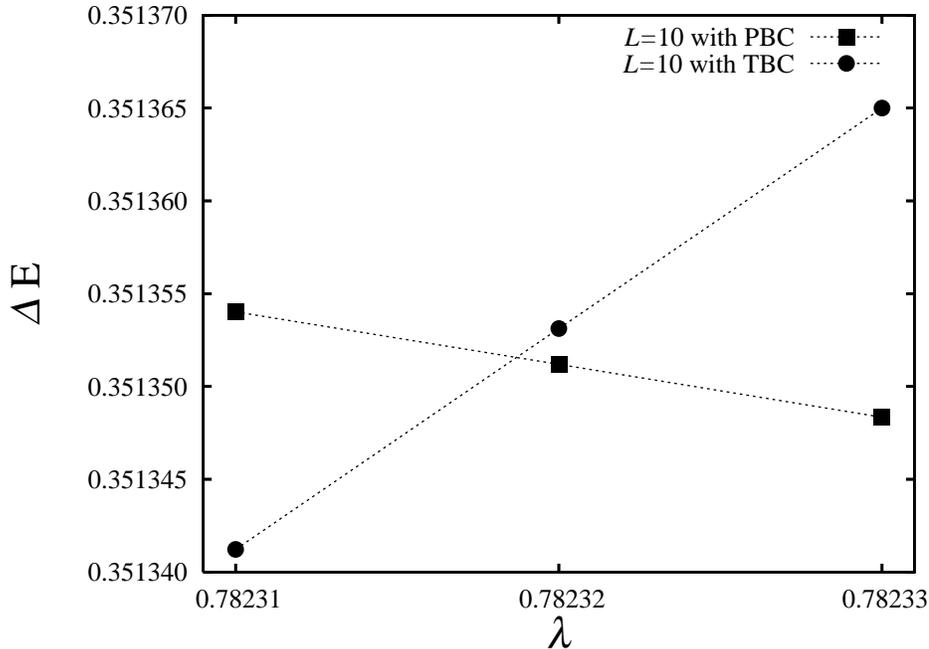} \caption{The
   crossing point for $L=10$ in lower $\lambda$\label{crossing_n10}}
  \end{figure}

\begin{figure}
   \includegraphics[scale=1.0] {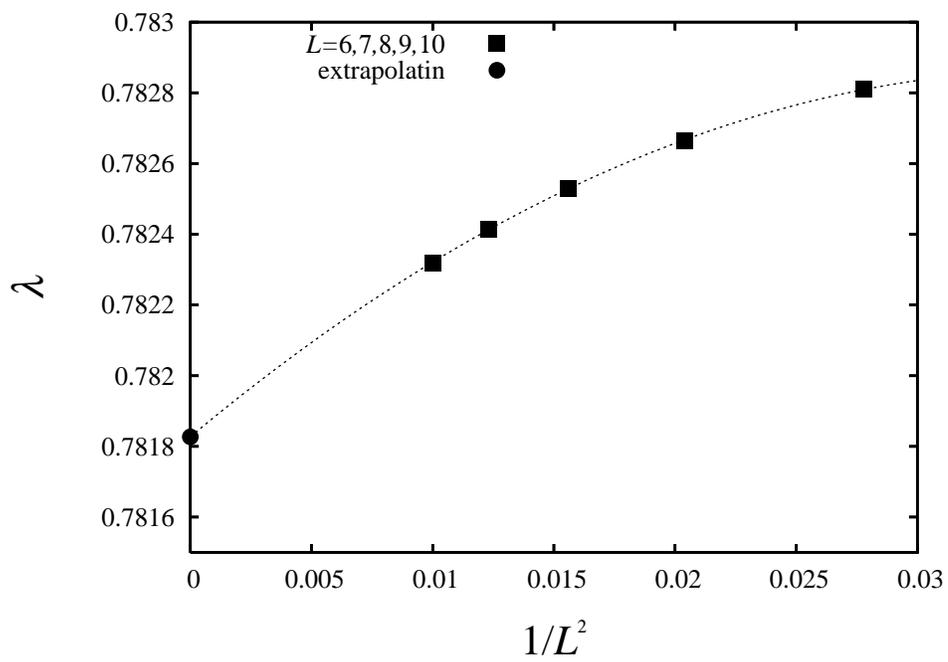}
   \caption{The lower BKT transition point. The extrapolated value is
   $\lambda = 0.78183$.}\label{low_tansp}
  \end{figure}

  Figure.$\ref{crossing_n10}$ shows the crossing point for $L=10$.  On
  the BKT transition point, we can eliminate logarithmic corrections
  including higher terms by the level spectroscopy, but still we have
  finite size corrections due to the irrelevant field $L_{-2}{\bar
  L}_{-2}{\bf 1}(x=4)$ \cite{cardy86:_operat_conten_of_two_dimen} which
  behaves order of $1/L^2, 1/L^4, {\rm etc}$. In
  figure.$\ref{low_tansp}$, we show the extrapolation of the level
  crossing as $L$ tends to infinity. The result is $\lambda = 0.78183$.
  
  \subsection{The BKT transition point in upper $\lambda$}
  The upper BKT critical point between the 6-fold degenerate ground
  state phase and the critical phase is given by the crossing point of
  the following scaling
  dimensions\cite{nomura95:_correl_gordon}\cite{otsuka05}
  \begin{eqnarray}
   \frac{9}{4} x_{3,0} &=& \frac{9}{4} \frac{L}{2\pi v} \Delta E(Q=3,
    PBC, C=1), \\
   x_{0,\frac{1}{2}} &=& \frac{L}{2\pi v} \Delta E(Q=0, TBC, C=1).
  \end{eqnarray}
  For $Q=0,3$, the charge conjugation $C$ can be a good quantum number
  as mentioned at the end of the subsection \ref{effective}.

\begin{figure}
   \includegraphics[scale=1.0]{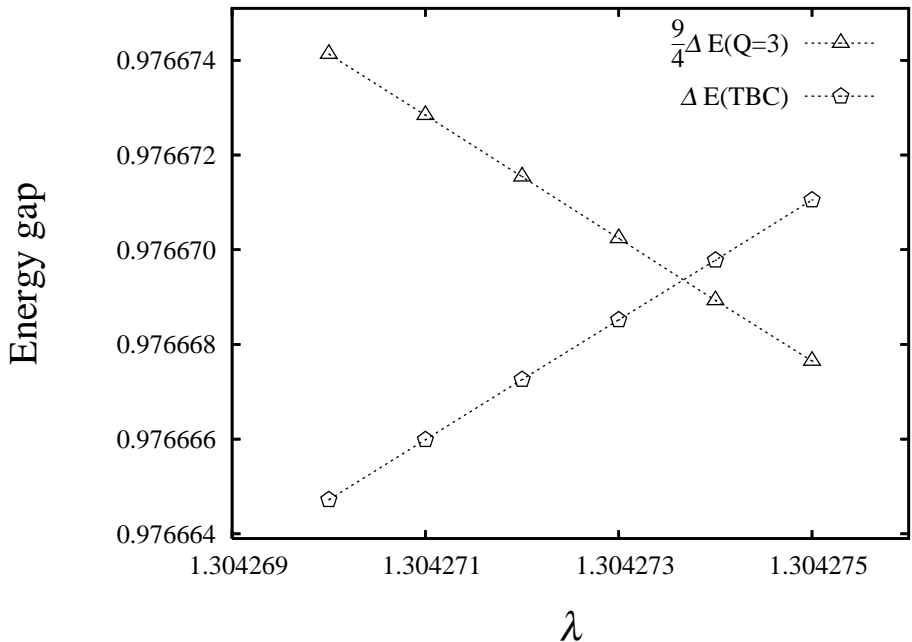}
   \caption{The crossing point for $L=10$ in upper
   $\lambda$}\label{up_cross10}
  \end{figure}

\begin{figure}
   \includegraphics[scale=1.0]{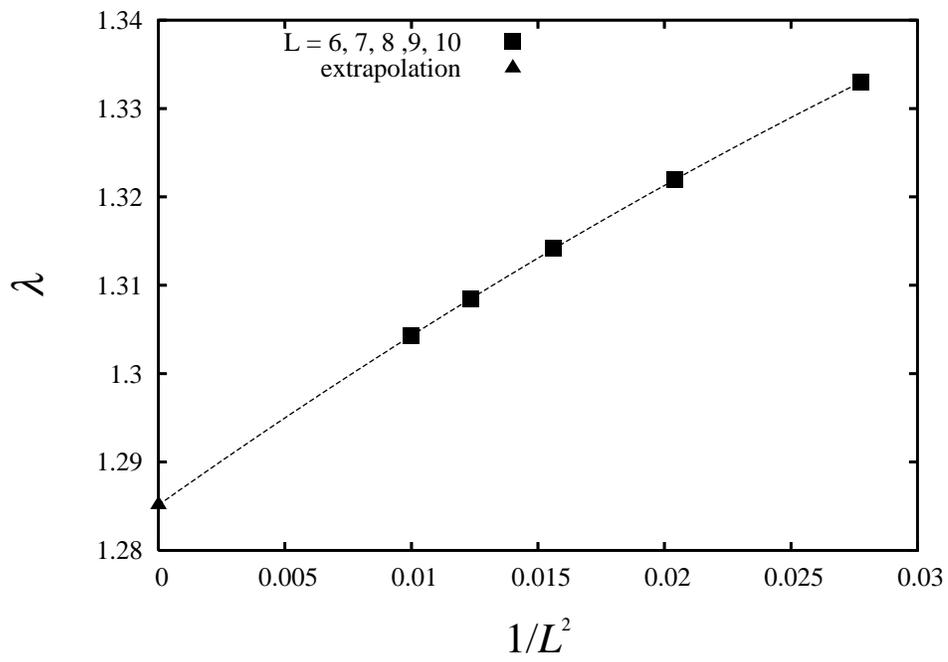} \caption{The
  upper BKT transition point. The extrapolated value is $\lambda =
  1.2851$.}\label{up_transp}
  \end{figure}

  The crossing point is shown in Figure.\ref{up_cross10} for
  $L=10$. This crossing eliminates the logarithmic correction order of
  $\mathcal{O}(1/\ln L)$. But still higher order corrections like
  $O(1/(\ln L)^2)$ can remain in this case. Anyway we extrapolate BKT
  transition point for large limit of the system size $L$ as in
  figure.$\ref{up_transp}$. We obtain $\lambda = 1.2851$.

    The product of $\lambda_{\mathrm{upper}}$ and $\lambda_{\mathrm{lower}}$
    can be used to check the consistency of the self duality; that is,
    $\lambda_{\rm upper} \cdot \lambda_{\rm lower}$ should be
    unity. Actually, from the numerical data, we get
    \begin{eqnarray}
     \lambda_{\rm upper} \cdot \lambda_{\rm lower} &=& 1.0047.
    \end{eqnarray}

  \subsection{Universality class}
   \subsubsection{Duality}
   \begin{figure}
    \includegraphics[scale=1.0]{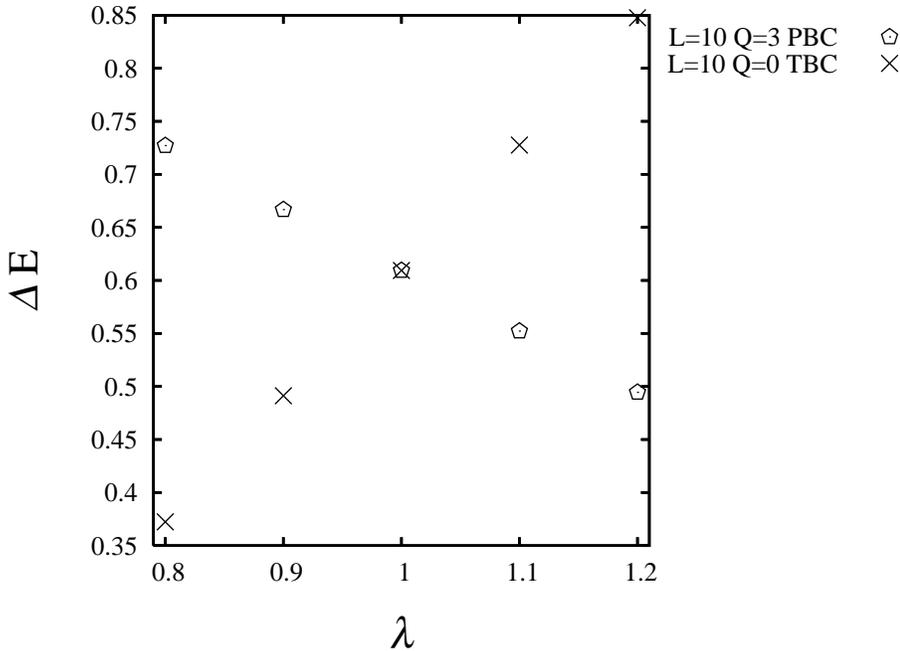}
    \caption{Level crossing occurs exactly at the self-dual point
    $\lambda = 1$.  These levels are $\Delta E(Q=3,{\rm PBC}, C=1)$ and
    $\Delta E(Q=0,{\rm TBC}, C=1)$ for $L=10$. The other system sizes
    also have the exact crossing.}  \label{fig:nocorrection}
   \end{figure}
   We numerically calculate the change of $K$ via $\lambda$ by
   the ratio of scaling dimensions:
   \begin{eqnarray}
    \frac{\Delta E(Q=3, {\rm PBC}, C=1)}{\Delta E(Q=0, {\rm TBC}, C=1)}
     = \frac{x_{3,0}}{x_{0,\frac{1}{2}}} = 36/K^2.
   \end{eqnarray}
   This ratio has the logarithmic correction $\mathcal{O} (1/(\ln L)^2)$
   in the vicinity of the BKT transition points. But at the self-dual point,
   $\Delta E(Q=3, {\rm PBC}, C=1)$ and $\Delta E(Q=0, {\rm TBC}, C=1)$
   exactly degenerate; therefore exactly $K=6$.  This is because the
   dual sine-Gordon model has higher symmetry at the self-dual point as
   was discussed in $(\ref{sG_dual})$,
   $(\ref{eq:dual_sineGordon})$. We show this in
   figure.$\ref{fig:nocorrection}$. The energy excitations: $\Delta E(Q=3,{\rm
   PBC}, C=1)$ and $\Delta E(Q=0, {\rm TBC}, C=1)$ exactly cross at
   $\lambda = 1$.

   von Gehlen and Rittenberg\cite{gehlen85:_z_n_symmet_quant_chain}
   pointed out that the spectra of the $Z_p$-symmetric quantum chains at
   the self-dual point are $p$-fold degenerated. Our novel point is not
   only the degeneracy but the connection of the discrete $Z_p$ model
   and the $Z_p$ dual sine-Gordon model.
   
   Even though when one study some other lattice model which doesn't
   have the self-duality explicitly, but if it could be mapped to the
   dual sine-Gordon model, one can check the self-dual point of the dual
   sine-Gordon model through this level crossing. For example, such an
   approach will be useful in the vicinity of the multicritical point
   studied by Otsuka \etal\cite{otsuka05}.

   \subsubsection{Ratio of scaling dimensions}
   \begin{figure}
     \includegraphics[scale=1.0]{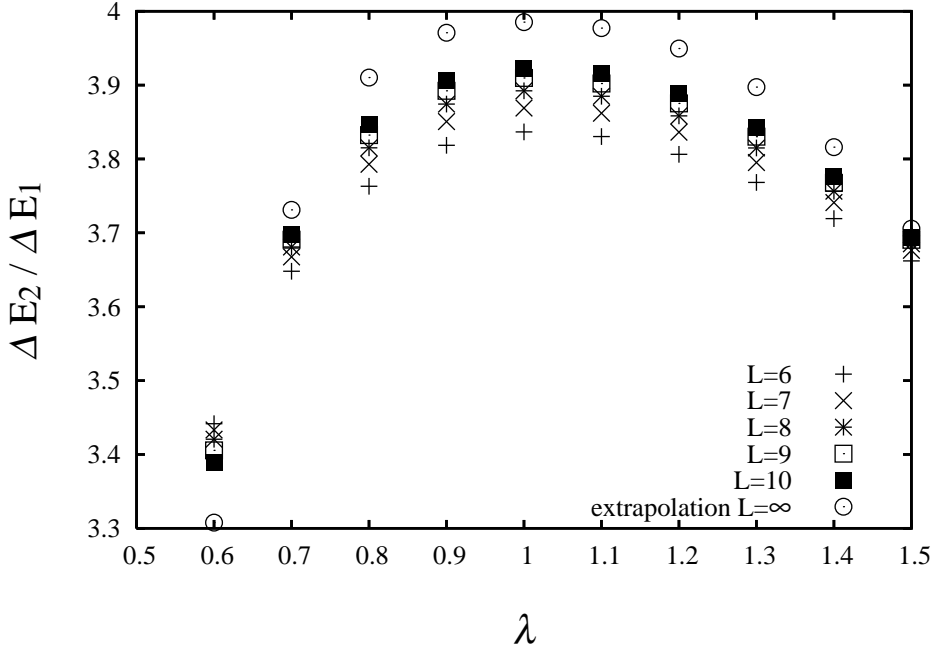}
     \caption{Ratio of scaling dimensions
     $x_{2,0}/x_{1,0}$. In the critical region the ratio
     $x_{2,0}/x_{1,0} \simeq 4$}\label{ratio4}
    \end{figure}
   The ratio of scaling dimensions
   \begin{eqnarray}
    \frac{x_{2,0}}{x_{1,0}}
     = \frac{\Delta E(Q=2, PBC)}{\Delta E(Q=1,PBC)} = 4 + \mathcal(L^{-2}) 
     + \mathrm{higher order}
   \end{eqnarray}
   is also useful to check the universality class of the BKT critical
   region. The finite correction behaves order of
   $\mathcal{O}(L^{-2})$. Figure.$\ref{ratio4}$ shows the ratio.
   \subsubsection{Central charge}
   Generally, in a BKT critical region a renormalization equation
   flows to the Gaussian fixed line. Therefore the universality class of
   the BKT critical region agrees with the Gaussian model. The Gaussian
   model is known to have the central charge $c=1$, so the BKT critical
   region is also characterized by $c=1$\cite{haldane81:_luttin}.

   From the conformal field theory the central charge is related to the
   ground state of the finite system as\cite{belavin84:_infin}
   \begin{eqnarray}
    E_{g}(L) = e_g L - \frac{\pi vc}{6L} \left( 
     1 + \mathcal{O}\left(\frac{1}{(\ln L)^3} \right) \right)
   \end{eqnarray}
   where $e_g$ is a free energy per site and $v$ is a velocity of the
   system. This is one of the important result of the conformal field
   theory.  In this case the logarithmic correction is small enough, so
   the central charge is numerically a good index of universality class.
   \begin{figure}
    \includegraphics[scale=1.0]{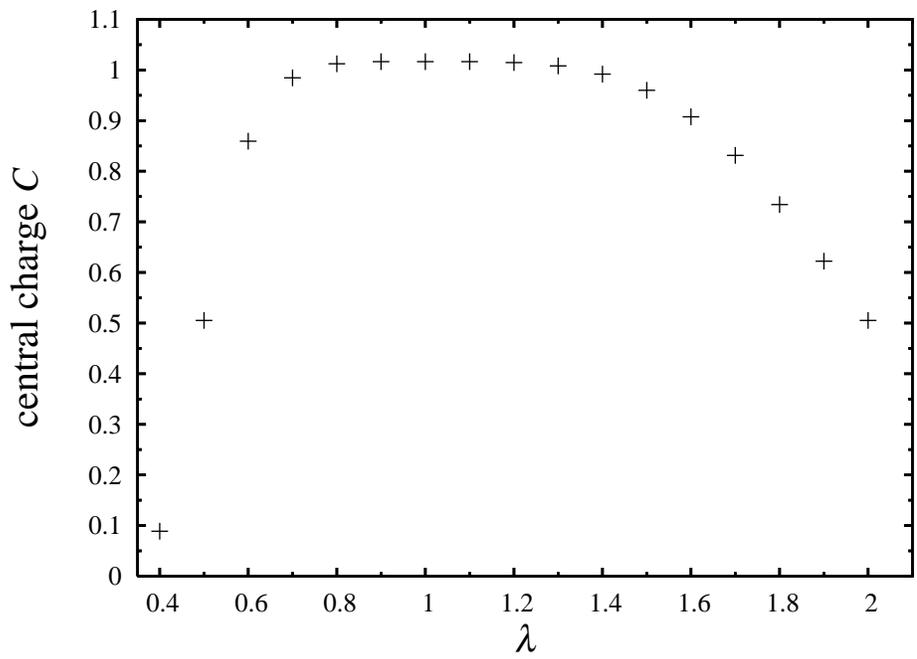}
    \caption{The central charge extrapolated from
    the finite system's results.}\label{fig:ccharge} 
   \end{figure}

   We calculate the effective central charge for the system sizes
   up to $L=10$ and extrapolated to $L\rightarrow\infty$. In Figure
   \ref{fig:ccharge}, we show the extrapolated values of the central
   charge. This result is consistent with the two transition points we
   have determined.

\section{Conclusion}
We extended the level spectroscopy to be able to determine the BKT
critical point between the multi-fold degenerated phase and the critical
region. For a physical application we studied the 1D quantum $p$-clock
model which has the $Z_p$ symmetry. The numerical calculations were
performed for $p=6$. And also we discussed about the self-duality of the
dual sine-Gordon model. In the section III-C, the self-dual point could
be numerically determined without logarithmic corrections. This is
because the level crossing of $\Delta E(Q=p/2,\mathrm{PBC})$ and $\Delta
E(Q=0,\mathrm{TBC})$ is the exact result from the
duality$(\ref{eq:dualtrans})$. There is no correction term in the langage
of the dual sine-Gordon model (\ref{sGeq1}). On the other hand, the BKT
transition point that we determined have some correction beause we
ignore irrelevant terms to derive $(\ref{sGeq2})$ and
$(\ref{sGeq3})$ from $(\ref{sGeq1})$. Our present approach will
be useful for the other models which need more highly accurate
calculations, for instance the cross-over near the multicritical point
dealed in \cite{otsuka05}.
\bibliographystyle{unsrt}
\bibliography{pclock}

\begin{thebibliography}{10}

\bibitem{berezinskii70:_destr}
Berezinskii~Z L.
\newblock {\em Zh. Eksp. Teor. Fiz.}, 61:1144, 1970.

\bibitem{j.m.kosterlitz73:_order}
Kosterlitz~J M and Thouless~D J.
\newblock {\em J. Phys. C: Solid State Phys}, 6:1181, 1973.

\bibitem{j.v.jose77:_renor}
Jos\'{e}~J V, Kadanoff~L P, Kirkpatrick S, and Nelson~D R.
\newblock {\em Phys.Rev.B}, 16:1217, 1977.

\bibitem{s.79:_phase_abelian}
Elitzur S, Pearson~R B, and Shigemitsu J.
\newblock {\em Phys. Rev. B}, 19:3698, 1979.

\bibitem{nomura94:_critic_s_xxz}
Nomura K and Okamoto K.
\newblock {\em J. Phys A: Math. Gen.}, 27:5773, 1994.

\bibitem{nomura95:_correl_gordon}
Nomura K.
\newblock {\em J. Phys A: Math. Gen.}, 28:5451, 1995.

\bibitem{nomura98:_su_z_bkt}
Nomura K and Kitazawa A.
\newblock {\em J. Phys A: Math. Gen.}, 31:7341, 1998.

\bibitem{tonegawa04:_magnet_s}
Tonegawa T, Okamoto K, Okunishi K, Nomura K, and Kaburagi M.
\newblock {\em Physica B}, 346:50, 2004.

\bibitem{h05:_level_spect_of_squre_lattic}
Otsuka H, Mori K, Okabe Y, and Nomura K.
\newblock {\em Phys. Rev. E}, 72:046103, 2005.

\bibitem{solyom81:_renor_hamil_potts}
S\'olyom J and Pfeuty P.
\newblock {\em Phys. Rev. B}, 24:218, 1981.

\bibitem{destri89:_twist_bound_condit_in_confor_invar_theor}
Destri C.
\newblock {\em Phys. Rev. B}, 223:365, 1989.

\bibitem{wiegmann78:_one_fermi}
Wiegmann~P B.
\newblock {\em J. Phys. C: Solid state Phys.}, 11:1583, 1978.

\bibitem{cardy86:_operat_conten_of_two_dimen}
Cardy~J L.
\newblock {\em Nucl. Phys. B}, 270:186, 1986.

\bibitem{affleck85:_univer_term_free_energ_critic}
Affleck I.
\newblock {\em Phys. Rev. Lett.}, 56:746, 1985.

\bibitem{gehlen85:_z_n_symmet_quant_chain}
von Gehlen~G and Rittenberg V.
\newblock {\em Nucl. Phys. B}, 257:351, 1985.

\bibitem{haldane81:_luttin}
Haldane F~D M.
\newblock {\em J. Phys. C: Solid state Phys.}, 14:2585, 1981.

\bibitem{belavin84:_infin}
Belavin~A A, Polyakov~A M, and Zamolodchikov~A B.
\newblock {\em Nucl. Phys. B}, B241:333, 1984.

\end{thebibliography}

\end{document}